\documentclass[letterpaper, 10 pt, conference]{ieeeconf}  

\IEEEoverridecommandlockouts                              

\usepackage{graphics} 
\usepackage{epsfig} 
\usepackage{mathptmx} 
\usepackage{times} 
\usepackage{amsmath} 
\usepackage{amssymb}  

\title{\LARGE \bf
Attention GhostUNet++: Enhanced Segmentation of Adipose Tissue and Liver in CT Images
}

\author{Mansoor Hayat$^{1}$, Supavadee Aramvith$^{2}$, Subrata Bhattacharjee$^{3}$   and Nouman Ahmad$^{4}$
\thanks{ This research is funded by Thailand Science Research and Innovation Fund Chulalongkorn University (IND\_FF\_68\_280\_2100\_039).}
}

\begin{document}

\maketitle
\thispagestyle{empty}
\pagestyle{empty}

\begin{abstract}

Accurate segmentation of abdominal adipose tissue, including subcutaneous (SAT) and visceral adipose tissue (VAT), along with liver segmentation, is essential for understanding body composition and associated health risks such as type 2 diabetes and cardiovascular disease. This study proposes Attention GhostUNet++, a novel deep learning model incorporating Channel, Spatial, and Depth Attention mechanisms into the Ghost UNet++ bottleneck for automated, precise segmentation. Evaluated on the AATTCT-IDS and LiTS datasets, the model achieved Dice coefficients of 0.9430 for VAT, 0.9639 for SAT, and 0.9652 for liver segmentation, surpassing baseline models. Despite minor limitations in boundary detail segmentation, the proposed model significantly enhances feature refinement, contextual understanding, and computational efficiency, offering a robust solution for body composition analysis. The implementation of the proposed \textbf{Attention GhostUNet++} model is available at: 
 {https://github.com/MansoorHayat777/Attention-GhostUNetPlusPlus}.
\newline

\indent \textit{Clinical relevance}— The \textbf{Attention GhostUNet++} model offers a significant advancement in the automated segmentation of adipose tissue and liver regions from CT images. Accurate delineation of visceral and subcutaneous adipose tissue, alongside liver structures, is critical for clinicians managing cardiometabolic disorders, including type 2 diabetes and cardiovascular diseases. By reducing reliance on manual annotations, the model enhances efficiency and scalability, paving the way for its integration into routine clinical workflows and large-scale body composition studies.
\end{abstract}

\section{Introduction}
Obesity is a significant risk factor for cardiometabolic diseases, including type 2 diabetes (T2D), cardiovascular disease (CVD), non-alcoholic fatty liver disease, and hypertension \cite{b1} \cite{b2}. Body composition (BC) analysis focuses on the distribution of fatty and non-fatty tissues, particularly in depots such as adipose tissue, muscle, liver, and bone, playing a crucial role in predicting and preventing these diseases  \cite{b3}. Among adipose tissue compartments, visceral adipose tissue (VAT) and subcutaneous adipose tissue (SAT) are key. VAT, located within the abdominal cavity \cite{b4, b5}. Similarly, SAT, located beneath the skin  \cite{b3}  \cite{b6, b7, b8, b17}.

This study introduces Attention GhostUNet++, an advanced deep learning model that incorporates Channel, Spatial, and Depth Attention mechanisms into the Ghost UNet++ bottleneck for automated segmentation of VAT, SAT, and liver regions from CT images. Using the AATTCT-IDS \cite{b10} and LiTS \cite{b11} datasets, the model achieved Dice coefficients of 0.9430 (VAT), 0.9639 (SAT), and 0.9652 (liver), outperforming baseline models in most metrics. However, challenges remain in SAT and VAT boundary segmentation, where Jaccard indices of 0.9639 (SAT) and 0.9430 (VAT) fell short of UNet’s performance in specific cases.

The proposed model reduces reliance on manual annotation, providing an efficient, scalable solution for BC analysis. It addresses challenges in large-scale imaging studies, paving the way for broader applications in multi-class segmentation and personalized healthcare, with future work focusing on boundary refinement and dataset generalization.
\begin{figure*}[t]
\centering
\includegraphics[width=\textwidth]{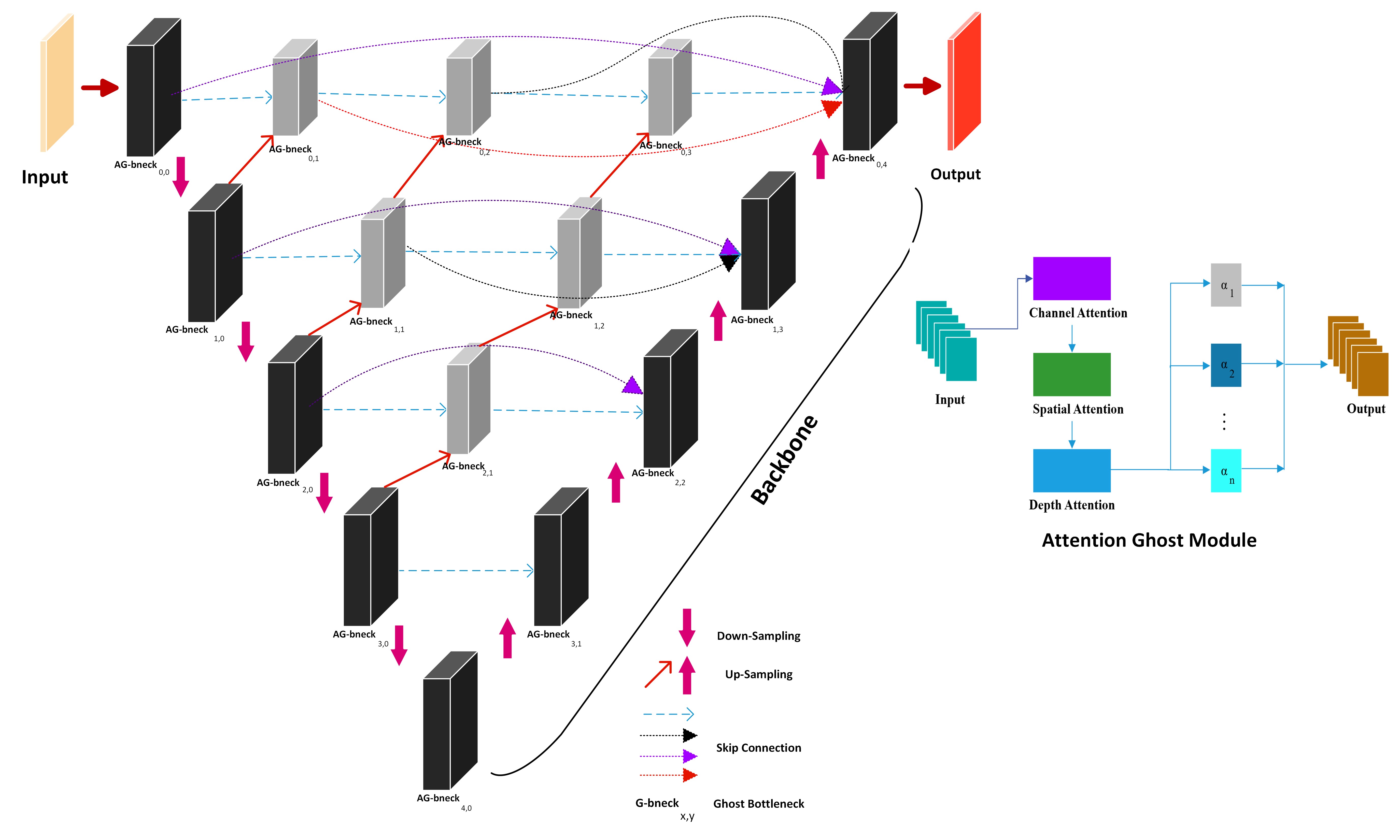}
\caption{\textbf{Attention GhostUNet++ Architecture.}}
\label{fig}
\end{figure*}

\section*{Methodology}
\subsection*{Ghost Module}
The Ghost-UNet \cite{b9} architecture is designed to enhance computational efficiency by minimizing redundancy in feature maps. Traditional convolution operations are computationally intensive for generating high-dimensional feature maps. The Ghost module addresses this limitation by employing lightweight operations to generate \textit{ghost feature maps}, which are lower-resolution representations of the input features. This significantly improves efficiency without sacrificing accuracy.

The Ghost module is mathematically expressed as:
\[
\mathbf{F}_o = \gamma (\mathbf{W} * \mathbf{F}_i) + \beta
\]
where:
\begin{itemize}
    \item \(\mathbf{F}_o\): High-resolution output feature map,
    \item \(\mathbf{F}_i\): Low-resolution ghost feature map,
    \item \(\mathbf{W}\): Weight tensor of the ghost convolution layer,
    \item \(\gamma\) and \(\beta\): Learnable scale and shift parameters,
    \item \(*\): Ghost convolution operation.
\end{itemize}

The ghost convolution operation is further defined as:
\[
\mathbf{F}_i[m] = \sum_{n} W[n, m] * \mathbf{X}[n] + b
\]
where \(m\) represents the ghost channels, \(n\) is the input feature map index, and \(b\) denotes the bias term.

\subsection*{Attention GhostUNet++ Architecture}
In the proposed \textbf{Attention GhostUNet++}, Ghost bottleneck layers are integrated into the UNet++ architecture, each enhanced with \textit{Channel, Spatial, and Depth Attention} mechanisms. These mechanisms improve feature refinement by dynamically emphasizing relevant regions in the feature maps while suppressing redundant information. The network consists of 15 bottleneck layers arranged within a nested architecture that supports robust contraction and expansion paths:
\begin{itemize}
    \item \textbf{Contraction Path}: Extracts features at multiple resolutions.
    \item \textbf{Expansion Path}: Reconstructs feature maps for precise segmentation.
\end{itemize}

Each bottleneck layer in the network can be represented as:
\[
\mathbf{F}_o = \mathcal{A}(\mathcal{G}(\mathbf{F}_i, \Theta))
\]
where:
\begin{itemize}
    \item \(\mathbf{F}_i\): Input tensor,
    \item \(\mathbf{F}_o\): Output tensor,
    \item \(\mathcal{G}\): Ghost bottleneck layer,
    \item \(\mathcal{A}\): Attention mechanism,
    \item \(\Theta\): Set of learnable parameters.
\end{itemize}

At each hierarchical level \(l\) in the network, the feature maps are calculated as:
\[
\mathbf{F}_l = \mathcal{P}(\mathcal{G}(\mathbf{F}_{l-1})) + \mathcal{U}(\mathcal{G}(\mathbf{F}_{l+1}))
\]
where:
\begin{itemize}
    \item \(\mathcal{P}\): Pooling operation,
    \item \(\mathcal{U}\): Up-sampling operation,
    \item \(\mathcal{G}\): Ghost module.
\end{itemize}

The final output of the network is:
\[
\mathbf{F}_{\text{final}} = \mathbf{F}_1 + \mathbf{F}_2 + \cdots + \mathbf{F}_n
\]
where \(n\) is the number of hierarchical levels in the architecture.
\begin{figure*}[t!]
\centering
\includegraphics[width=\textwidth]{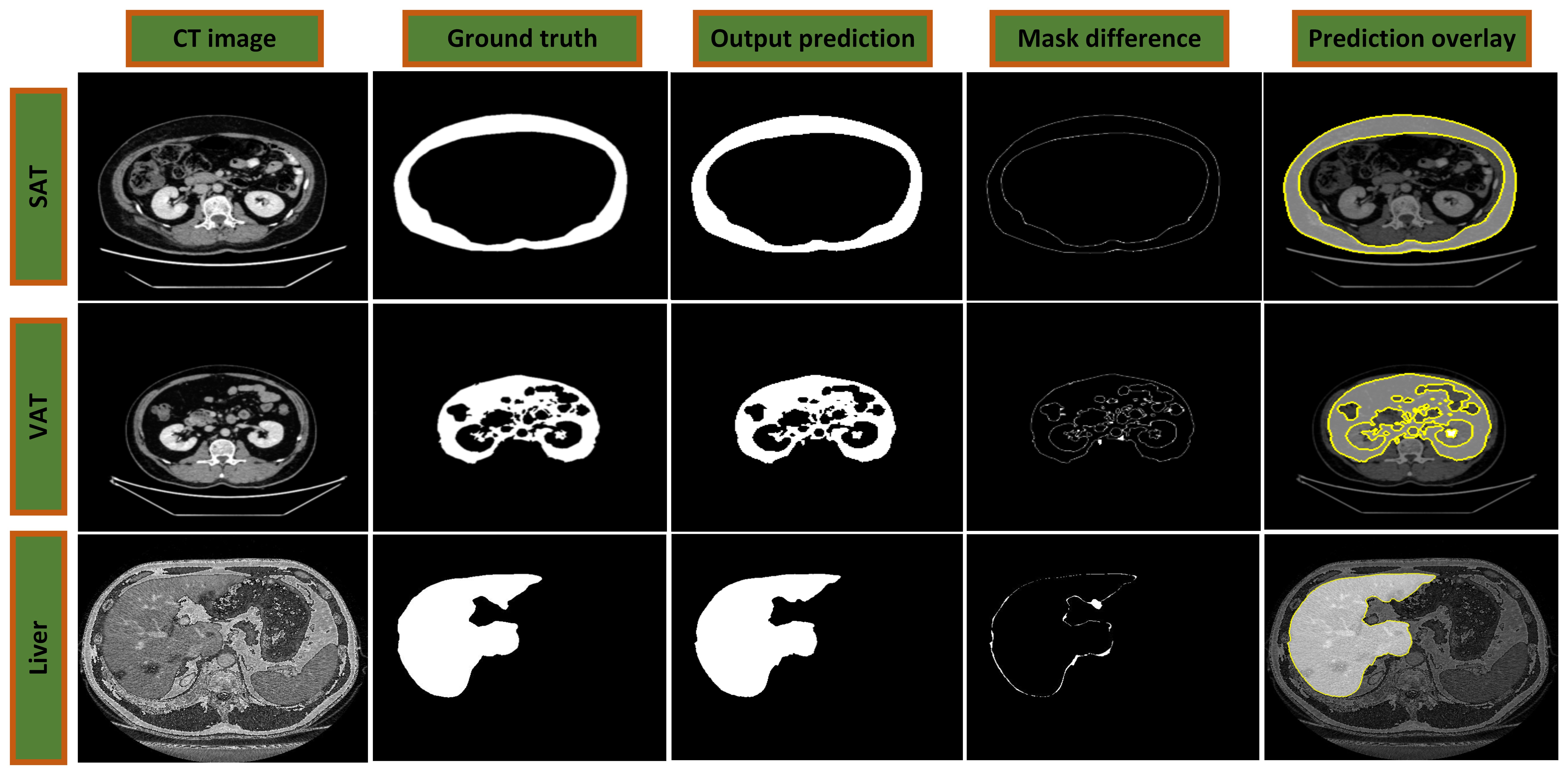}
\caption{\textbf{Segmentation results for randomly selected CT image examples, the columns represent (1) the original CT image, (2) ground truth annotations, (3) model-predicted segmentation output, (4) mask differences between ground truth and predictions, (5) predicted segmentation masks overlaid on the original CT image.}}
\label{fig}
\end{figure*}
This innovative integration of Ghost bottleneck layers with attention mechanisms enables \textbf{Attention GhostUNet++} to achieve high accuracy, computational efficiency, and improved feature refinement, making it a robust solution for medical image segmentation.

\section{Experimental Results}
\subsection*{Datasets}
\subsection*{Datasets and Preprocessing}
This study evaluates the proposed \textbf{Attention GhostUNet++} model using two datasets: the \textit{Abdominal Adipose Tissue CT Image Dataset (AATTCT-IDS)} \cite{b10} and the \textit{Liver Tumor Segmentation Benchmark (LiTS)} \cite{b11}. These datasets provide annotated CT images for segmenting SAT, VAT, and liver regions.

The \textbf{AATTCT-IDS} dataset \cite{b10} includes 13,732 CT slices (3,213 annotated) from 300 subjects, focusing on SAT and VAT. Challenges arise from individual variability and overlapping boundaries between these compartments. The \textbf{LiTS} dataset \cite{b11} contains 201 CT volumes annotated for liver regions, featuring diverse liver shapes and pathologies. Liver segmentation was prioritized in this study.

Preprocessing included resizing CT slices to a consistent resolution, normalizing pixel intensities, and cropping volumes to focus on regions of interest. Data augmentation (e.g., rotations, flipping, scaling) was applied to enhance variability and reduce overfitting.

These datasets were critical for validating the robustness and generalizability of the \textbf{Attention GhostUNet++} model in segmenting complex anatomical structures across varying scenarios.

\subsection*{Training Settings}
The proposed \textbf{Attention GhostUNet++} model was implemented in PyTorch 2.0 and trained on an Nvidia 3090Ti GPU. Xavier initialization ensured effective weight scaling. The datasets were split into training, validation, and test sets (70:20:10), with data augmentation (rotations, flipping, scaling, and intensity variations) applied to improve generalization.

Training used the Adam optimizer \cite{b16} (\(1 \times 10^{-4}\) initial learning rate, cosine annealing decay) and a combined Dice and cross-entropy loss. Mini-batches of size 16 optimized GPU usage, while early stopping (100-epoch patience) prevented overfitting. The best-performing model was selected based on the validation Dice coefficient (DC).

Segmentation accuracy was evaluated using the DC and Jaccard index (JI):
\[
\text{Dice Coefficient} = \frac{2|\mathbf{P} \cap \mathbf{G}|}{|\mathbf{P}| + |\mathbf{G}|} = \frac{2 \sum_{i} P_i G_i}{\sum_{i} P_i + \sum_{i} G_i}
\]
\[
\text{Jaccard Index} = \frac{|\mathbf{P} \cap \mathbf{G}|}{|\mathbf{P} \cup \mathbf{G}|} = \frac{\sum_{i} P_i G_i}{\sum_{i} P_i + \sum_{i} G_i - \sum_{i} P_i G_i}
\]
where \(P_i\) and \(G_i\) represent the predicted and ground truth segmentation at pixel \(i\), respectively.

\subsection*{Performance Evaluation}
The performance of the proposed \textbf{Attention GhostUNet++} model was compared against baseline models (UNet \cite{b12}, UNet++ \cite{b13}, ResUNet \cite{b14} and GhostUNet++ \cite{b15} for VAT, SAT, and liver. Evaluation metrics included the DC and JI, which assess segmentation accuracy and overlap, respectively.

The proposed model outperformed the baseline models in most cases. As indicated in Table I, for VAT segmentation, it achieved a DC of 0.9430 and a JI of 0.9430, closely competing with UNet’s JI of 0.9491. In SAT segmentation, it achieved a DC of 0.9639, matching UNet’s JI of 0.9807 but indicating room for improvement in handling boundary details. For liver segmentation, Attention GhostUNet++ demonstrated superior performance with a Dice coefficient of 0.9652 and a JI of 0.9496, outperforming all baselines.

\begin{table}[h!]
\centering
\caption{Mean Segmentation Dice and Jaccard Scores for Different Targets in AATTCT-IDS \cite{b10} and LiTS \cite{b11} Datasets}
\label{tab:results}
\begin{tabular}{|l|l|c|c|c|}
\hline
\textbf{Method}          & \textbf{Metrics}       & \textbf{VAT} & \textbf{SAT} & \textbf{Liver} \\ \hline
UNet \cite{b12}                     & Dice coefficient       & 0.9057       & 0.9604       & 0.8746         \\ 
                         & Jaccard index          & \textbf{0.9491} & \textbf{0.9807} & 0.8456      \\ \hline
UNet++ \cite{b13}                    & Dice coefficient       & 0.8742       & 0.8741       & 0.9468         \\ 
                         & Jaccard index          & 0.8157       & 0.8639       & 0.9311         \\ \hline
ResUNet \cite{b14}                   & Dice coefficient       & 0.9184       & 0.9482       & 0.9587         \\ 
                         & Jaccard index          & 0.9021       & 0.9653       & 0.9412         \\ \hline
GhostUNet++ \cite{b15}                & Dice coefficient       & 0.8847       & 0.8916       & 0.9554         \\ 
                         & Jaccard index          & 0.7916       & 0.8451       & 0.9318         \\ \hline
Attention GhostUNet++   & Dice coefficient       & \textbf{0.9430} & \textbf{0.9639} & \textbf{0.9652} \\ 
                         & Jaccard index          & 0.9430       & 0.9639       & \textbf{0.9496} \\ \hline
\end{tabular}
\end{table}

These results confirm the effectiveness of the Attention GhostUNet++ \cite{b15} model for medical image segmentation. While achieving state-of-the-art performance across most tasks, minor limitations in SAT segmentation suggest opportunities for further refinement to enhance boundary accuracy and overall generalizability.

Fig. 2 showcases example segmentation outputs from the test dataset. Ground truth annotations are displayed alongside predictions from each model. The Attention GhostUNet++ \cite{b15} model exhibits superior boundary adherence and accurate region segmentation, particularly in challenging cases where SAT and VAT boundaries overlap or are difficult to distinguish. Additionally, for liver segmentation, the proposed model effectively captures the organ's irregular contours, delivering results that clearly outperform those of the baseline models.

These visual comparisons underscore the model's capability to handle complex segmentation tasks with enhanced precision and boundary accuracy, making it a robust solution for medical imaging applications.

\section{Conclusion}
We proposed \textbf{Attention GhostUNet++}, a novel deep learning architecture for segmenting SAT, VAT, and liver regions in CT images. By integrating \textit{Channel, Spatial, and Depth Attention} mechanisms into Ghost-Net bottleneck layers, the model achieves enhanced feature refinement and contextual understanding with computational efficiency.

Experiments on AATTCT-IDS\cite{b10} and LiTS\cite{b11} datasets demonstrated state-of-the-art performance, with DCs of 0.9430 (VAT), 0.9639 (SAT), and 0.9652 (liver). Visual comparisons highlighted the model’s ability to accurately segment complex anatomical structures, reducing boundary errors and outperforming baseline models.

This automated solution reduces reliance on manual annotations and enhances scalability for clinical and research applications. Future work will focus on addressing limitations, extending to multi-class tasks, and validating across diverse datasets, paving the way for efficient and accurate medical imaging tools.

\section*{Limitations and Future Work}
The proposed \textbf{Attention GhostUNet++} model faces minor limitations in handling fine-grained boundaries, particularly for SAT segmentation, and its generalizability to diverse imaging modalities remains untested. It also requires further evaluation on datasets with significant variations in anatomical structures and imaging quality. Future work will focus on enhancing boundary detection, expanding validation to multi-class tasks and diverse datasets, and optimizing computational efficiency for clinical deployment. Additionally, integrating advanced techniques, such as edge-aware attention, could improve performance in challenging regions. Extending the model to segment complex pathological structures will further enhance its utility in diagnostic and therapeutic applications.


\begin{thebibliography}{00}
\bibitem{b1} Mokdad, Ali H., Earl S. Ford, Barbara A. Bowman, William H. Dietz, Frank Vinicor, Virginia S. Bales, and James S. Marks. "Prevalence of obesity, diabetes, and obesity-related health risk factors, 2001." Jama 289, no. 1 (2003): 76-79.
\bibitem{b2} Kaess, Bernhard M., Jacek Jozwiak, Miroslaw Mastej, Witold Lukas, Wladyslaw Grzeszczak, Adam Windak, Wieslawa Piwowarska et al. "Association between anthropometric obesity measures and coronary artery disease: a cross-sectional survey of 16 657 subjects from 444 Polish cities." Heart 96, no. 2 (2010): 131-135.
\bibitem{b3} Kullberg, Joel, Anders Hedström, John Brandberg, Robin Strand, Lars Johansson, Göran Bergström, and Håkan Ahlström. "Automated analysis of liver fat, muscle and adipose tissue distribution from CT suitable for large-scale studies." Scientific reports 7, no. 1 (2017): 10425.
\bibitem{b4} Tanaka, Muhei, Hiroshi Okada, Yoshitaka Hashimoto, Muneaki Kumagai, Hiromi Nishimura, and Michiaki Fukui. "Distinct associations of intraperitoneal and retroperitoneal visceral adipose tissues with metabolic syndrome and its components." Clinical Nutrition 40, no. 5 (2021): 3479-3484..
\bibitem{b5} Tanaka, M., Okada, H., Hashimoto, Y., Kumagai, M., Nishimura, H., \& Fukui, M. (2020). Intraperitoneal, but not retroperitoneal, visceral adipose tissue is associated with diabetes mellitus: a cross-sectional, retrospective pilot analysis. Diabetology \& Metabolic Syndrome, 12, 1-10.
\bibitem{b6} Christen, T., Sheikine, Y., Rocha, V. Z., Hurwitz, S., Goldfine, A. B., Di Carli, M., \& Libby, P. (2010). Increased glucose uptake in visceral versus subcutaneous adipose tissue revealed by PET imaging. JACC: Cardiovascular Imaging, 3(8), 843-851.
\bibitem{b7} Kelley, D. E., Thaete, F. L., Troost, F., Huwe, T., \& Goodpaster, B. H. (2000). Subdivisions of subcutaneous abdominal adipose tissue and insulin resistance. American Journal of Physiology-Endocrinology and Metabolism, 278(5), E941-E948.
\bibitem{b8} Smith, S. R., Lovejoy, J. C., Greenway, F., Ryan, D., deJonge, L., de la Bretonne, J., ... \& Bray, G. A. (2001). Contributions of total body fat, abdominal subcutaneous adipose tissue compartments, and visceral adipose tissue to the metabolic complications of obesity. Metabolism-Clinical and Experimental, 50(4), 425-435.
\bibitem{b9} Kazerouni, I. A., Dooly, G., \& Toal, D. (2021). Ghost-UNet: an asymmetric encoder-decoder architecture for semantic segmentation from scratch. IEEE Access, 9, 97457-97465.
\bibitem{b10} Ma, Z., Li, C., Du, T., Zhang, L., Tang, D., Ma, D., ... \& Sun, H. (2024). AATCT-IDS: A benchmark Abdominal Adipose Tissue CT Image Dataset for image denoising, semantic segmentation, and radiomics evaluation. Computers in Biology and Medicine, 177, 108628.
\bibitem{b11} Bilic, P., Christ, P., Li, H.B., Vorontsov, E., Ben-Cohen, A., Kaissis, G., Szeskin, A., Jacobs, C., Mamani, G.E.H., Chartrand, G. and Lohöfer, F., 2023. The liver tumor segmentation benchmark (lits). Medical Image Analysis, 84, p.102680.
\bibitem{b12} Ronneberger, O., Fischer, P., \& Brox, T. (2015). U-net: Convolutional networks for biomedical image segmentation. In Medical image computing and computer-assisted intervention–MICCAI 2015: 18th international conference, Munich, Germany, October 5-9, 2015, proceedings, part III 18 (pp. 234-241). Springer International Publishing.
\bibitem{b13} Zhou, Z., Rahman Siddiquee, M. M., Tajbakhsh, N., \& Liang, J. (2018). Unet++: A nested u-net architecture for medical image segmentation. In Deep Learning in Medical Image Analysis and Multimodal Learning for Clinical Decision Support: 4th International Workshop, DLMIA 2018, and 8th International Workshop, ML-CDS 2018, Held in Conjunction with MICCAI 2018, Granada, Spain, September 20, 2018, Proceedings 4 (pp. 3-11). Springer International Publishing.
\bibitem{b14} Rahman, H., Bukht, T. F. N., Imran, A., Tariq, J., Tu, S., \& Alzahrani, A. (2022). A deep learning approach for liver and tumor segmentation in CT images using ResUNet. Bioengineering, 9(8), 368.
\bibitem{b15} Ahmad, N., Strand, R., Sparresäter, B., Tarai, S., Lundström, E., Bergström, G., Ahlström, H. and Kullberg, J., 2023. Automatic segmentation of large-scale CT image datasets for detailed body composition analysis. BMC bioinformatics, 24(1), p.346.
\bibitem{b16} Bock, S. and Weiß, M., 2019, July. A proof of local convergence for the Adam optimizer. In 2019 international joint conference on neural networks (IJCNN) (pp. 1-8). IEEE.
\bibitem{b17} Hayat, Mansoor. "Squeeze \& Excitation joint with Combined Channel and Spatial Attention for Pathology Image Super-Resolution." Franklin Open 8 (2024): 100170.
\end{thebibliography}
\end{document}